# An enhanced performance for H.265/SHVC based on combined AEGBM3D filter and back-propagation neural network

**L. Balaji**
Department of ECE, Velammal Institute of Technology, Chennai, India
E-mail: maildhanabal@gmail.com

**K. K. Thyagharajan**
ECE Department, RMD Engineering College, Chennai, India
E-mail: kkthyagharajan@yahoo.com

**Abstract**
This paper deals with the latest video coding standard H.265/ SHVC, a scalable extension to High Efficiency Video Coding (HEVC). HEVC introduces new coding tools compared to its predecessor and is backward compatible with all types of electronic gadgets. The gadgets with different display capabilities cannot be offered the same quality video due to the constraints in transmission bandwidth is a major problem. One solution to this problem will be the compression of the video sequence which is focused in this paper to preserve or increase PSNR while reducing bit-rate besides a novel method implemented in SHVC encoder. The novel method undergoes a combined AEGBM3D (adaptive edge guided block-matching and 3D) filtering and back-propagation technique. The technique includes an AEGBM3D filter which avoids spatial redundancy and de-noise frames; hence enhancement in PSNR is achieved. The obtained PSNR of the video is compared with the set threshold PSNR to maintain PSNR above the threshold by repeated AEGBM3D filtering. The BP technique based on the neural network machine learning approach continually restrains the output if the input block does not contain a feature they were trained to recognize. This frequent control over the output produces few bits; hence reduction in bit-rate is achieved. The simulation results show that the proposed technique delivers an average increment of 0.16 and 0.25dB in PSNR and an average decrement of 28 and 37% in bit-rate for ×1.5 and ×2 spatial ratios, respectively, compared with the existing methods.

## 1 Introduction

A new video compression standard is introduced by ITUT VCEG and ISO/IEC MPEG in 2013 popularly known as H.265/High Efficiency Video Coding (HEVC). HEVC encoder surpasses its predecessor H.264/Advanced Video Coding (AVC) by higher compression, i.e., up to 45.54% in terms of bit-rate [1]. Such high compression is achieved by its sophisticated number of intra-modes and viable inter-predicting modes. Owing to this increased number of modes and its flexibility for compression, the computation complexity of HEVC is more than AVC.

The scalable extension of HEVC had major attention from the industry to call upon developing a standard. Later in 2014, MPEG and ITU have launched the extended version of HEVC, known as SHVC [2]. The scalable extension uses inter-layer predicting modes along with the superior coding features of HEVC. Having noticed, HEVC is more complex, its extended version of scalable achievement is likely to be extensively more complex than its predecessor H.264/Scalable Video Coding (SVC). As a result, one major challenge in the general implementation of SHVC is reducing its computational complexity. The computational complexity in SHVC increases due to one of the major factor is mode search from the available number of modes with minimum rate-distortion cost. In addition to mode search, more computation part includes partitioning of its Coding Tree Unit (CTU). Although many algorithms were proposed for its predecessor H.264/SVC, to reduce the computational complexity, none of the algorithms will be suitable for SHVC. Since SHVC uses several modes and flexible inter-prediction modes, the algorithms called for reducing the computation complexity does not exactly suitable for SHVC.

Every frame in HEVC is partitioned into square-shaped equal-sized blocks. The fundamental block is the code tree block (CTB). For every CTB of luminance, two CTBs of related chrominance samples do exist. The entire subset of one luminance CTB, two chrominance CTB and their associated rule among these blocks are known as coding tree unit (CTU). Unlike H.264/AVC, HEVC employs a quad-tree-based coding structure that can support various sizes of a block. To enhance more compression, every CTU can split into squares of smaller blocks known as coding units (CUs). Like the CTU, every CU block contains one block of luminance and two blocks of chrominance samples and their associated rule needed for transformation. To achieve improved coding efficiency, every CU is split into 4 small CUs. The procedure of splitting CTU into CUs is lengthened for F iterations. The F iterations denote the highest CU depth in the CTU quad-tree structure. Afterward, the CUs can be split into small prediction units (PUs) and transform units (TUs) [3–5].

HEVC offers intra-prediction modes more in number to minimize spatial redundancy and more elastic motion compensation to minimize temporal redundancy unlike H.264/AVC [1]. To compensate spatial redundancy, HEVC holds 35 luminance intra-prediction modes while 9 Intra modes in H.264/AVC. Furthermore, intra-prediction planned withvarioussizesofblocksrangefrom4×4to64×64while in H.246/AVC, it 4×4 to 16×16. The intra-prediction coding procedure, in HEVC grants 35 intra-modes altogether DC and Planar modes for the luminance component of every PU. The size of the PU [1] determines the maximum number of modes to be predicted as likely mode using the rate-distortion cost in HEVC.

Apart from intra-modes, the encoder decides among motion merge mode, skip mode and explicit encoding of motion information. For every PU block to be coded, the motion merge mode occupies generating a record of formerly coded spatial and temporal adjacent PUs recognized as candidates. The motion data for the present PU is duplicated from a chosen candidate, rather encoding a motion vector for the PU; as a replacement, only the index of a candidate in the motion merge record is encoded in addition to the residual. In skip mode, the encoder indicates the index of a motion merge candidate and the motion parameters for the present PU are duplicated from the chosen candidate, not including residual information to be sent. In explicit encoding mode, Symmetric and Asymmetric Motion Partitions (AMP) are done in inter-coded CUs. In AMPs, the CUs are partitioned into several small PUs of non squared shape size. AMPs can be done for CUs size from 64×64 downward to 16×16, succeeding in coding efficiency. Since, PUs depicts the conventional shape of objects, to be more accurate, which we do not require to split further [1]. Every inter-predicted PU contains a record of motion parameters, which includes a motion vector, a reference frame index and a reference list flag.

The scalable version of HEVC (SHVC) exploits the sophisticated prediction methods like HEVC. To get better coding efficiency [6], SHVC incorporates inter-layer motion and inter-layer texture prediction. In inter-layer motion prediction, the motion data (motion vector and index of reference frame) of the co-located CU in the base layer are appended to the merge candidate list [7] adding together the adjacent PUs presently used in HEVC. In inter-layer texture prediction, a reconstructed signal of a related CU in the base layer is interpolated. Followed by, a CU in the enhancement layer will be predicted based on the resulting signal. Here, interlayer reference frames are used as reference frames adding with temporal reference frames [2].

In the sight of the above, we prepared an attempt to enhance the performance of SHVC. It is imposed by a hybrid image blocking CAFBP scheme for PSNR improvement while compressing bits. AEGBM3D filter applied to avoid spatial redundancies and frame de-noising; hence improvement in PSNR is achieved. BP technique rooted from machine learning neural network is integrated to restrain the error signal, which reduces bit-rate. The rest of the paper is organized as follows: Sect. 2 explains the work related to SHVC, Sect. 3 explains the CAFBP scheme, the simulation results are discussed in Sect. 4, and finally, Sect. 5 concludes the scheme.

## 2 Related work

Lately, numerous works related to SHVC focused on low complexity computation, fast mode decision, rate control techniques, etc. Many computation complexity reduction methods have been already discussed for HEVC [7–18]. In [7] a fast mode search algorithm was implemented for intra-prediction. A CTU splitting and prune methods are implemented near the beginning for intra-prediction [8]. Having considered, inter-prediction is a more complex section in HEVC with low delay and random access profile, different methods discussed to reduce the complexity in inter-prediction [9–16]. Merge mode identification is recommended near the beginning which utilizes the root block mode, all-zero block mode and the motion estimation is discussed in [12]. With the use of motion deviation, the size of the coding unit (CU) can be selected quickly is suggested [14]. A successful CU size detection scheme is discussed in [15] which undergo two strategies to reduce the computation. One strategy is estimating the quad-tree depth range. The second strategy minimizes the computation of motion estimation for all small block sizes. From the time when SHVC utilizes the superior coding features of HEVC, the discussed computation complexity reduction schemes for HEVC can be applied for the computation complexity reduction of both the layers (base and enhancement layer).

Lately, many computation complexities minimizing techniques have been practiced for SHVC encoder [19–25]. A mode search scheme near the beginning termination [20] and a dynamic search range scheme is discussed for quality scalability [19]. The mode search scheme uses the rate-distortion cost values of adjacent blocks to expect the rate-distortion cost of the block to be coded in the higher layer (enhancement layer).Thedynamicsearchrangeschemeuses the motion information of the lower layer (base layer) to dynamically alter the search range in the higher layer (enhancement layer). A fast mode method based on Bayesian classifier for quality scalability is discussed [21], which exploits adjacent blocks mode and related block in the base layer to expect the best likely mode for the present block in the enhancement layer. While a dynamic scheme is discussed for spatial scalability which expects the search range data for the enhancement layer blocks exploiting the motion information of related blocksinbaselayer[22]. Few works,[19–22]discussedatthe early start of SHVC performance validated based on HEVC reference encoder. Now, we intend to build up a computation complexity minimization scheme particularly planned for quality and spatial scalable versions of HEVC. With the proposed method the encoder complexity is minimized by avoiding checking all the CU sizes for the CTU partition.

[26]suggested a technique to empower differential coding in an adaptable codec plan without influencing the interior coding, therefore enabling a functional execution to reclaim single-layer equipment or programming segments. This is accomplished by making an extra reference picture called improved between layer reference and embeddings it to the enhancement layer (EL) decoded picture buffer and reference picture sets. Building up a plan for foreseeing the CTU structure for the scalable extension of HEVC in the spatial and quality scalability is focused [27]. The algorithm utilizes the CTU structure of the previously coded CTUs in the higher layers (EL) and lower layer (BL) to anticipate the size of coding units of the CTU blocks to-be-coded in the higher layer. Using two higher layers (EL) a reduced complexity method is proposed for SNR adaptive extension of HEVC [28]. The proposed method possibly can make encoding or broadcasting of a few diverse quality adaptations of similar video information in bit-stream an alluring recommendation for the information conveyance industry, taking into account financially savvy digital media conveyance to an assortment of playback show gadgets.

So far, all proposed methods may reduce the complexity of the encoder and by saving encoding time, but the expansion in information brought by several administrations requires more compressed data and transmits over networks. Besides, conveying such video information over systems needs a precise control of the bit-rate from coders to coordinate unbending limitations on transmission capacity and QoS. A few commitments have been proposed to together encode adaptable stream, yet without considering the effect of bitrate proportion among layers with compressed execution. In [29], the effect of the bit-rate proportion among layers on the encoder execution is first explored for a few UHD scalable methods proposed. To our extent, still, no method proposed for the scalable extension of HEVC which can enhance PSNR with lesser bits. The proposed hybrid block matching technique using AEGBM3D and BP technique which improves PSNR is described in Sect. 3, the subsection describes the impact of AEGBM3D filter, and the BP technique based on neural networks is explained in detail.

## 3 Combined AEGBM3D filter and BP scheme

In this proposed work, a combined AEGBM3D filter with BP technique-based machine learning neural network technique is implemented. The purpose of AEGBM3D filter is to avoid redundancies among the frame spatially, and besides, it can de-noise the frames; hence an improvement in PSNR is achieved. The BP technique based on the neural network machine learning approach restrains the error signal, to reduce bit-rate. In the following section, a detailed description of the AEGBM3D filter, BP technique and CAFBP scheme is presented.

### 3.1 AEGBM3D filter

AEGBM3D is a non-local image modeling based Fig. 1 on adaptive high-order group-wise models. A video sequence is split into frames, frames into macroblocks further split into blocks. The blocks are processed with a denoising algorithm to remove the redundant information and noise in the frame. One such video de-noising algorithm [31] executes grouping and filtering which removes redundancy while maintaining the original signal. This grouping and filtering are recalled in AEGBM3D filter so as to reduce the computation complexity by avoiding redundancy and for de-noising.

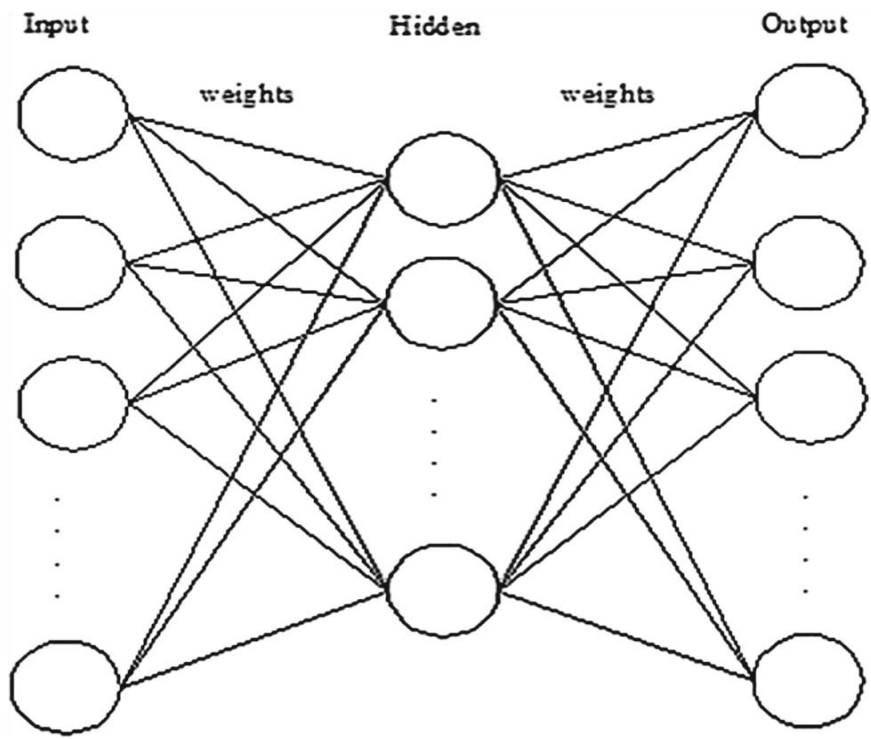

**Fig. 1** Machine learning model

Grouping is the process to identify the similar blocks. Similar 2D blocks are piled mutually to form 3D group of similar items is called as a group [30]. The filter process the blocks to de-noise the frames by executing a predictive search block matching method [31] along with two passes, respectively; the first pass undergoes block matching with hard thresholding, while the second pass undergoes block matching with wiener filtering. For every block grouping and filtering is performed to de-noise the frames of a video sequence. The grouping is accomplished by a predictive search block matching method that finds similar blocks among the reference and adjacent blocks, and their resemblance is measured in 3D field that extents both temporal and two spatial dimensions. By implementing this method the computation complexity is significantly reduced when compared to the full search method. Next, apply 3D transform on the group to generate a vastly sparse depiction of the original signal. The filtering (either hard thresholding or Weiner) is applied to the transform coefficients for reduction as well as noise removal while maintaining the important part of the original signal. Finally, the estimates of all groups are obtained by applying the inverse 3D transform.

A set of overlapped blockwise estimates are obtained for every block after executing grouping and filtering. The estimates are collected and are then combined by the weighted averaging process in which the weights are inversely proportional to the reduction in group spectrum during filtration and therefore freely reciprocal to the entire variance of every group.

Thus, two passes are applied to de-noise the frames in the video, first-pass perform grouping and hard thresholding to generate a fundamental estimate (intermediary), followed by the second pass which performs grouping and wiener filtering which takes the spectrum from the fundamental estimates. This joint 3-D transform considerably improves the effectiveness of spectral image approximation.

### 3.2 Back-propagation technique

Figure 1 shows the generic model of a neural network, which contains input, hidden and output layers. The output layers are compared with the desired outputs so that a small error signal is expected with the training sets as shown in Fig. 2. One neural network that is useful in addressing such problems is the feed-forward network.

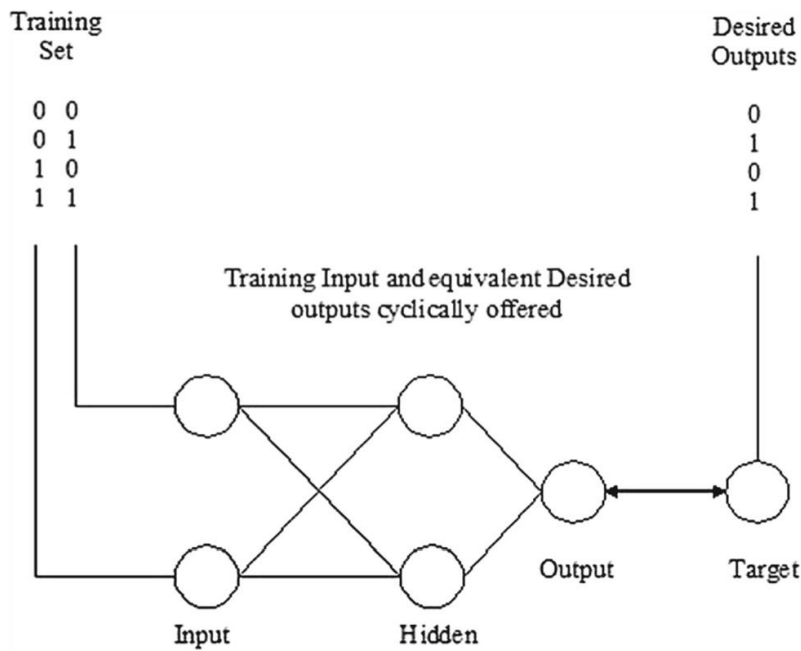

**Fig. 2** Training sets with desired outputs

(A) One of the set of **p** training input patterns is applied to the input layer.

$x_p = (x_{p1}, x_{p2}, ... x_{pn})$ which may be a binary or real numbered vector.

(B) The initiations of units in the hidden layer are calculated by taking their net input (the sum of the initiation of the input layer units they are connected to accumulate by their respective connection weights) and passing it through a transfer function.

(i) Summation of all inputs to the hidden layer unit $j$ to give net input of $j$

$$net_j = \sum_{i=1}^{n} W_{ji} X_i \quad \text{(when } j \text{ has multiple inputs)} \tag{1}$$

Squash using the activation function

$$oh_j = \frac{1}{1+e^{-net_j}} \tag{2}$$

All input passes through $j$ function.

(C) The initiation of the hidden layer units computed in (B) is then used in the initiation of the output units. Here all output of $j$ is made as input to $k$ while passing all output through the same transfer function.

(i) The net input to output unit $k$

$$net_k = \sum_{j=1}^{L} W_{kj}\, oh_j \tag{3}$$

(ii) Result of output unit $k$ (it is also calculated from the net input of $k$ that will pass through the same transfer function)

$$oo_j = \frac{1}{1+e^{-net_k}} \tag{4}$$

### 3.2.1 Backward pass

(A) The difference between the actual initiation of each output unit and the desired target initiation ($d_k$) for that unit is found, and this difference is used to generate an error signal for each output unit. A quantity called delta is then calculated for all output units.
(i) Error signal of actual output $oo_k$, and its desired output $d_k$

$$(d_k - oo_k) \tag{5}$$

(ii) Delta term which is a partial derivative of the nonlinear activation function for each output unit is equal to its error signal multiplied by the output of that unit multiplied by(1—its output) as

$$\delta_{ok} = (d_k - oo_k)oo_k(1 - oo_k) \tag{6}$$

(B) The error signals for the hidden layer units are then calculated by taking the sum of the deltas of the output units. A particular hidden unit connects is multiplied by the weight that connects the hidden and output unit. The deltas for each of the hidden layer units are then calculated. (i) Error signal for each hidden unit $j$.

$$\sum_{k=1}^{W} \delta_{o_k} W_{kj} \tag{7}$$

(ii) Delta term which is a partial derivative of nonlinear activation function for each hidden unit $j$ is equal to its error signal multiplied by its output, multiplied by (1—its output) as

$$\delta_{h_j} = (oh_j)(1 - oh_j) \sum_{k=1}^{W} \delta_{o_k} W_{kj} \tag{8}$$

(C) The weight error derivatives (WED) for every weight between the hidden and output units are ascertained byt aking the delta of every output unit and multiplying it by the initiation of the concealed unit it interfaces with. These weight error derivatives are then used to change the weights between the hidden and output layers.

$$wed_{jk} = \delta_{ok}(oh_j) \tag{9}$$

i.e., to calculate the weight error derivative between hidden unit $j$ and yield unit $k$ take the delta term of output unit k and increase it by the output (initiation) of hidden unit $j$.

(D) The weight error derivatives for every weight between the input unit $j$ and shrouded unit $j$ are computed by taking the delta of each hidden unit and multiplying it by the initiation of the input unit it interfaces with (i.e., that input pattern $x_i$). These weight error derivatives are then used to change the weights between the input and hidden layers.

$$wed_{ij} = \delta_{h_j}(x_i) \tag{10}$$

To change the actual weights themselves, a learning rate parameter n is used, which controls the sum of weights that areupdatedduringeveryBPcycle.Theweightsatonce($t$+1) between the hidden up and output layers are set using the weights at once and the weight error derivatives between the hidden and output layers using the following equation.

$$w_{jk}(t+1) = w_{jk}(t) + \eta(wed_{jk}) \tag{11}$$

Inasimilarwaytheweightsarechangedbetweentheinput and hidden units

$$w_{ij}(t+1) = w_{ij}(t) + \eta(wed_{ij}) \tag{12}$$

Using this method, every unit in the network receives an error signal that describes its relative contribution to the total error between the actual output and the target output. Based on the error signal received, the weights connecting the units in different layers are updated.

These two passes are repeated many times for different input patterns and their targets until the error between the actual output of the network and its objective yield is acceptably small for all of the members of the set of training inputs.

This type of training can be applied to much larger networks than the XOR network to solve much more complex problems, yet the fundamental two-pass cycle continues the same. As the network trains, units in the hidden layer organize themselves such that different units learn to recognize different features of the total input space. For example, if a network were trained to react to a pixel image of the letter 'T', one unit may create as an element finder for the vertical bar on the highest point of the 'T'. After training, when given a subjective new input pattern that is noisy or incomplete, the units in the hidden layer will react with an active output if the new input contains that resembles the feature of the individual units learned to recognize during training. On the other hand, hidden layer units tend to restrain their outputs if the input pattern does not contain a feature they were trained to recognize. These networks tend to create inward internal relationships between units to arrange the training data into classes of patterns. In this way, they develop an inner representation that enables them to the desired outputs when given the training inputs. This same interior representation can be applied to inputs that were not used during training. The BP network will group these new inputs as per the features they share with the training inputs, i.e., these networks can sum up. The main steps are as follows

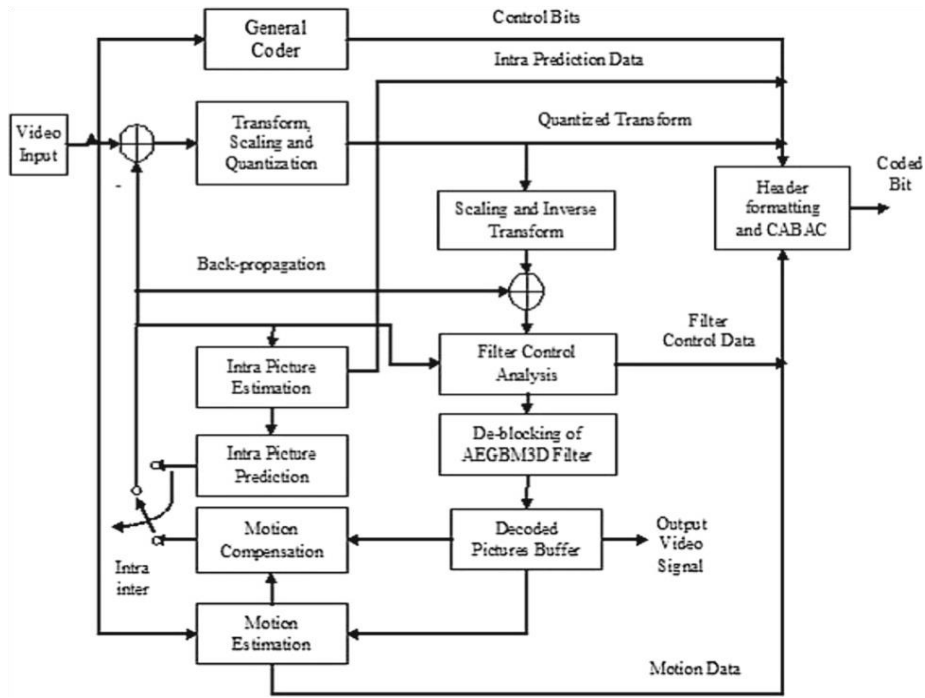

**Fig. 3** Proposed block of SHVC encoder using AEGBM3D filter

1. Initialize weights to small random values.
2. Select a training vector pair (input and the corresponding output) from the training set and present the input vector to the inputs of the network.
3. Calculate the actual outputs (forward pass).
4. According to the difference between actual and desired outputs (error signal). Adjust the weights $W_o$ and $W_h$ to reduce the difference (backward pass).
5. Repeat from step 2 for all training vectors.
6. Repeat from step 2 until the error is acceptably small.

### 3.3 CAFBP scheme

The proposed model of HEVC encoder architecture is shown in Fig. 3. The architecture obtains YUV video sequence which is converted into frames. The Frames are being converted into pixel blocks of variable sizes. Each block contains the extracted Luminance and Chrominance components of a frame and as per the variance of the extracted feature, the frameundergoblocksizesof8×8,16×16,32×32,64×64 blocks whichever is suitable. Then apply AEGBM3D filter which is specially designed for spatial predictive filtering techniques. It identifies and controls spatial redundancies that exist among frames by coding some original blocks via spatial prediction and other coding techniques. Check PSNR of a frame whether falls below the set threshold (highest PSNR among all frames in a video sequence). If PSNR is lower than the set threshold, apply the filtering technique until it upholds the PSNR above or equal to the set threshold. Followed by, the BP technique is applied to restrain the error signal, such that it reduces bit-rate, by continually controlling their outputs using SHVC encoding techniques.

Exploiting temporal relations that exist among blocks and subsequent frames, the changes between frames are encoded. This is accomplished via motion estimation and compensation where searches are performed on adjacent frames to form motion vectors that predict qualities of the target block. Identifyandtakeadvantageofanyremainingspatialredundancies that exist among frames by encoding only the deviations between original and predicted blocks through quantizing, transforming, and entropy coding. The steps involved in the proposed technique as follows

1. Extract all frames in a video sequence
2. Based on the variance of extracted feature from each frame, split into blocks of 8×8, 16×16, 32×32 and 64×64.
3. Apply AEGBM3D filter to avoid spatial redundancy and-noise the frame.
4. Check PSNR of a frame with the set threshold PSNR (extracted as highest PSNR among all frames in a video)
5. If PSNR falls above the threshold go to step 7 else go to step 6.
6. Apply filtering technique until it upholds PSNR above or equal to the set threshold, followed by, the BP technique restrains the error signal such that it reduces bit-rate.
7. Encode the frame using HEVC encoding techniques (quantization, transformation and entropy encoding).
8. Go to step 2.

## 4 Experimental results

The performance of the proposed scheme is evaluated with SHM reference software 12.1 [32]. In the implementation process, we tested 4 standard database video sequences (Kimono, Park scene, Duckstakeoff and Traffic) and 2 realtime video sequences (gate and home) are shown in Fig. 4 for quality and spatial scalability. We used the base layer and one enhancement layer for these two scalabilities. As per the SHM test conditions, spatial scalability is conducted for two ratios(×1.5and×2).Theproposedschemeiscomparedwith the previously proposed method [27] and the SHM reference software [32] (Fig. 5). As per the video resolution, the threshold PSNR will be calculated. The input video is split into frames. Frames are blocked into sub-frame or sub-imagesuchas64×64to8×8.

If the image has a higher resolution, higher-level blocking will be selected. Otherwise, it starts from 8×8. Every frame undergoes AEGBM3D filter. Then determine the PSNR of each frame. Select the highest PSNR among all frames in the entire video sequence. The highest PSNR is set as a threshold PSNR for the input video. Until the PSNR of input video reaches the threshold PSNR, the blocking of images will proceed from 8×8 to 64×64 which are controlled by the BP technique.

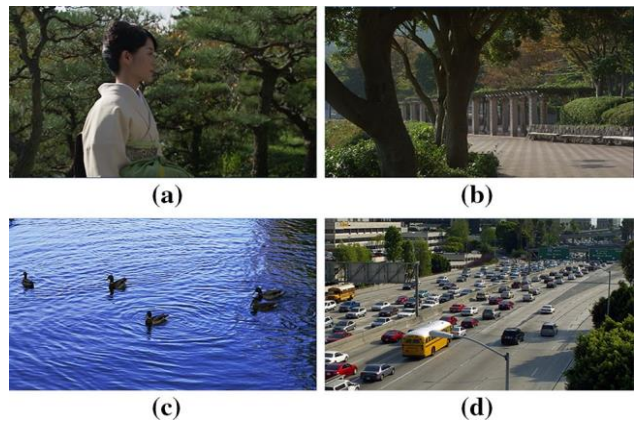

**Fig.4** Standardvideosequences.**a** Kimono,**b**Parkscene,**c** Duckstakeoff and **d** Traffic

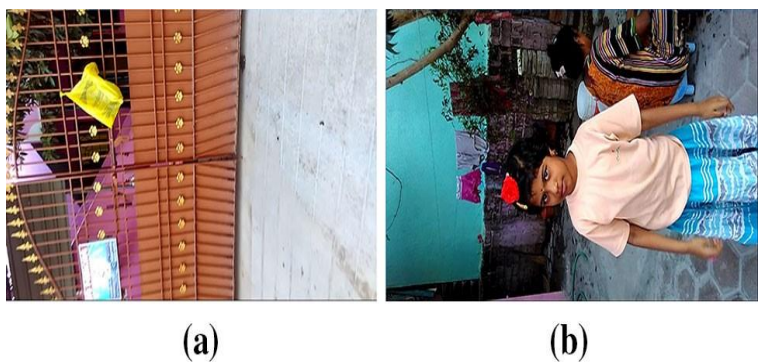

**Fig. 5** Real-time video sequences. **a** Gate and **b** home

In our test, we compare random access quality (RA (Q)), random access spatial (RA(S)), low-delay P (LDP) and low delay B (LDB) profiles of SHVC. For Quality scalability, we set the quantization parameters of the base layer and enhancement layer as (26, 22), (30, 26), (34, 30) and (38, 34). In Spatial scalability, the quantization parameters are set as (22, 22),(26,26),(30,30)and(34,34) for base layer and enhancement layer in both spatial ratios (×1.5 and ×2). For class B video sequence both spatial ratio's conducted, while for class A type of video sequence one spatial ratio (×2) alone conducted.

Table1illustratesthecombinedscalabilityunderfourprofiles such as RA (Q), RA (S), LDP and LDB with an average increase of 0.16 dB in PSNR and an average decrease of 28% in bit-rate under 1.5× spatial resolution. In other cases, an averageincreaseof0.25dBinPSNRandanaveragedecrease of 0.37% in bit-rate under 2× spatial resolution in comparison with the previously proposed method.

Figures 6, 7, 8 and 9 show the rate-distortion (RD) curves for different video sequences under SNR and spatial scalabilities. The RD curve plots the data between bit-rate versus PSNR of each sequence in all configurations such as random access—SNR, ×1.5, ×2; low-delay P—×1.5, ×2; and low-delay B—×1.5, ×2. It is also compared with the standard SHM 12.1 reference encoder for SHVC [32] and HM 6.1 reference encoder for HEVC single-layer coding along with the previously proposed method [27]. Although, single-layer HEVC coder outperforms scalable-level encoding techniques in terms of bit-rate and PSNR in some video sequence, when it comes to scalability, single-layer coding cannot provide such flexibility. Comparing our proposed scheme with the existing methods, the proposed scheme does better in terms of PSNR and bit-rate, except a slight increase in encoding time.ThisencodingtimeisduetotheBPscheme which repeats to maintain the error signal to the minimum and achieves to reduce the number of bits.

## 5 Conclusion

This paper provides an overview of the latest scalable coding standard SHVC. SHVC adopts a scalable coding architecture; with the machine learning approach of BP is implemented. Efficient BP is achieved by inter-layer reference picture processing modules. In contrast to the previous scalable coding standard SVC, the EL codec in SHVC can be built by repurposing existing single-layer HEVC codec cores. We compare the random access (RA), low-delay P (LDP) and low delay B (LDB) profiles of H.265/SHVC. For Quality scalability, we use the quantization parameters of the base layer and enhancement layer as (26, 22), (30, 26), (34, 30) and (38, 34). In Spatial scalability, the quantization parameters are (22, 22), (26,26),(30,30)and(34,34) for base layer and enhancement layer in both spatial ratios (×1.5 and ×2). For class B video sequence both spatial ratios are conducted, while for class A type of video sequence spatial ratio (×2) alone conducted. In all cases by examining we conclude that the combination of AEGBM3D and BP gives an average increment of 0.16 dB and 0.25 dB in PSNR and an average decrement of 28 and 37%inbit-ratefor×1.5and×2spatialratios,respectively.In our future work, we concentrate on the adaptable PSNR threshold selection and it works as per the resolution of the video sequence, the PSNR will be selected.

Table 1 Quality and spatial scalability in all 4 configurations [RA (Q), RA (S), LDP and LDB]

| Profile | Layer | [27] versus proposed | | | | | [32] versus proposed | | | | |
|---|---|---|---|---|---|---|---|---|---|---|---|
| | | PSNR(dB) | | | | BR (%) | PSNR(dB) | | | | BR (%) |
| | | *YUV* | *Y* | *U* | *V* | | *YUV* | *Y* | *U* | *V* | |
| RA (Q) | BL | 0.37 | 0.37 | 0.41 | 0.42 | −0.29 | 0.02 | 0.07 | −0.08 | 0.08 | −0.03 |
| | EL | 0.62 | 0.60 | 0.64 | 0.66 | −0.55 | 0.14 | 0.39 | −0.88 | −0.96 | −0.07 |
| RA (S) | 1.5 | 0.12 | 0.11 | 0.12 | 0.13 | −0.48 | 0.17 | 0.29 | −0.52 | 0.28 | −0.07 |
| | 2 | 0.21 | 0.20 | 0.21 | 0.22 | −0.55 | 0.18 | 0.19 | −0.53 | 0.07 | −0.03 |
| LDP | 1.5 | 0.07 | 0.07 | 0.07 | 0.08 | −0.16 | 0.21 | 0.18 | −0.81 | 0.27 | −0.02 |
| | 2 | 0.06 | 0.06 | 0.06 | 0.06 | −0.16 | 0.25 | 0.06 | −0.43 | −0.13 | −0.02 |
| LDB | 1.5 | 0.06 | 0.06 | 0.06 | 0.07 | −0.18 | 0.19 | 0.07 | −0.81 | −0.12 | −0.04 |
| | 2 | 0.12 | 0.12 | 0.13 | 0.13 | −0.23 | 0.29 | 0.42 | 0.58 | 0.41 | −0.04 |
| Average | 1.5 | 0.16 | 0.15 | 0.16 | 0.17 | −0.28 | 0.15 | 0.15 | −0.55 | 0.13 | −0.04 |
| | 2 | 0.25 | 0.25 | 0.26 | 0.27 | −0.37 | 0.22 | 0.27 | −0.31 | −0.15 | −0.04 |

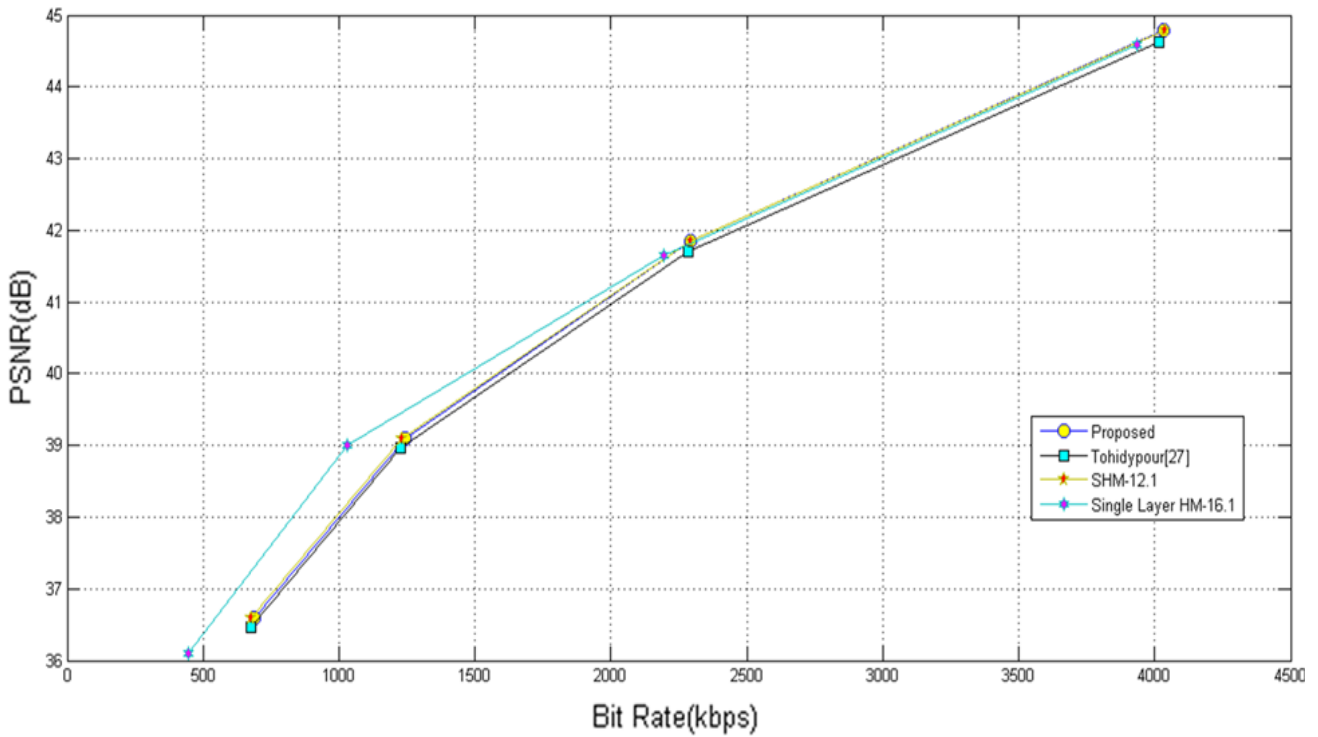

**Fig. 6** RD curve for RA ×1.5 gate sequence

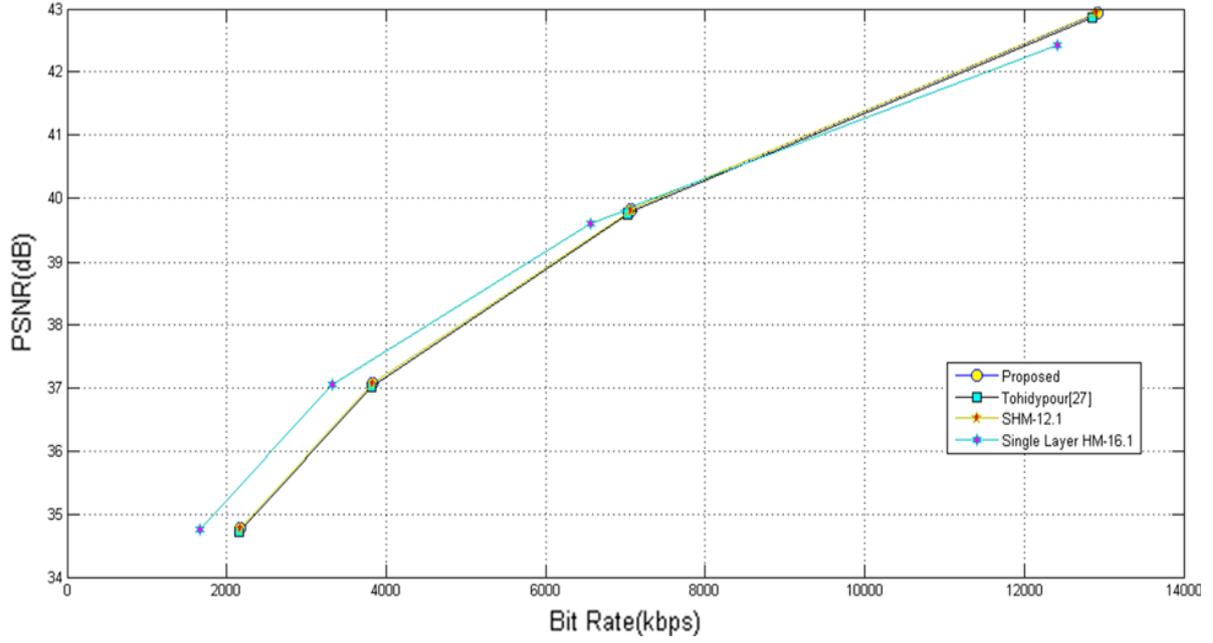

**Fig. 7** RD curve for LDP ×2 home sequence

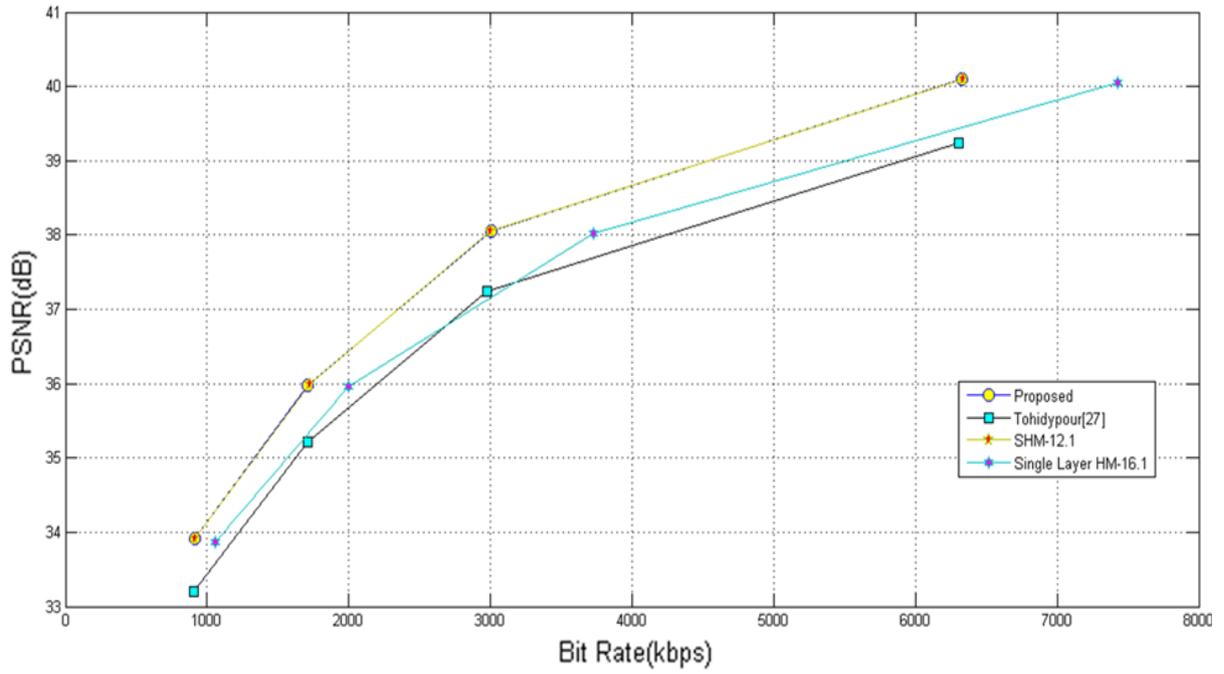

**Fig. 8** RD curve for RA SNR Kimono sequence

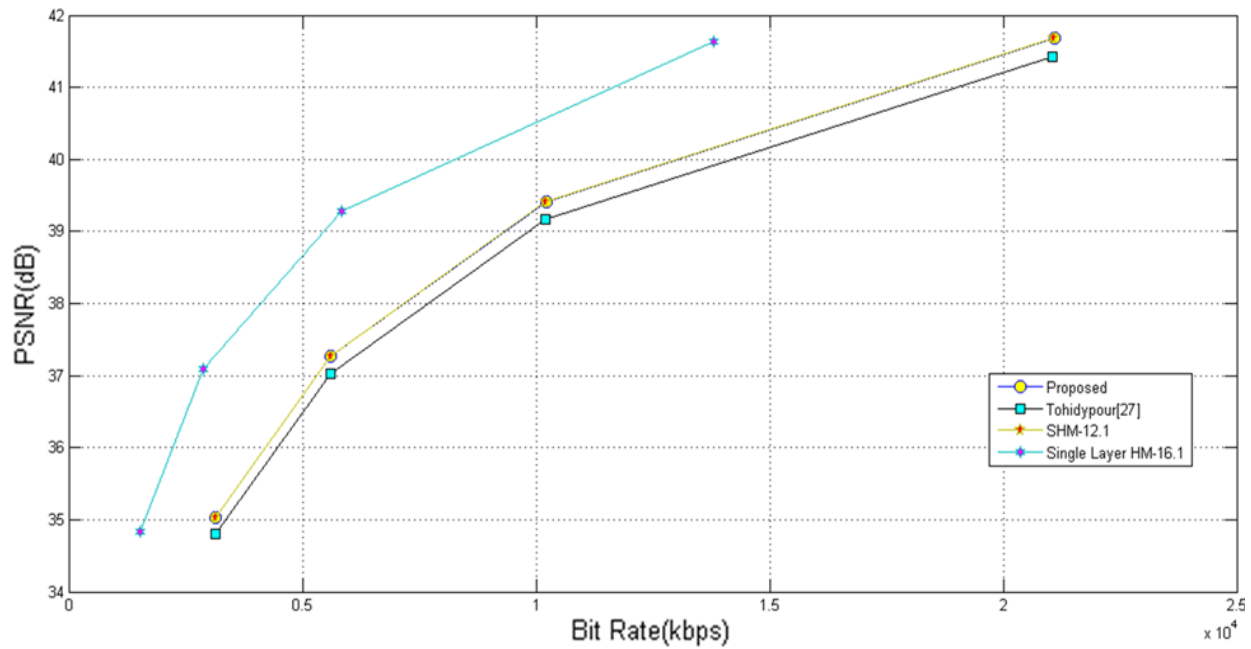

Fig. 9 **RD curve for LDB ×2 traffic sequence**